\documentclass[10pt]{article}

\usepackage[margin=1in]{geometry}
\usepackage{textcomp}
\usepackage{parskip}
\usepackage{amsmath}
\usepackage{authblk}
\usepackage[backend=bibtex, style=alphabetic, sorting=ynt]{biblatex}
\usepackage{graphicx}
\usepackage{wrapfig}
\usepackage{float}
\usepackage{caption}
\usepackage{hyperref}

\graphicspath{{./images/}}

\begin{document}

\title{Developing a Bubble Chamber Particle Discriminator Using Semi-Supervised Learning}
\author[1]{B.~Matusch\thanks{Corresponding: \url{brendon.matusch@snolab.ca} or \url{analysis@picoexperiment.com}}}
\affil[1]{SNOLAB, Lively, Ontario, P3Y 1N2, Canada}

\author[2]{C.~Amole}
\affil[2]{Department of Physics, Queen's University, Kingston, K7L 3N6, Canada}

\author[3]{M.~Ardid}
\affil{Departament de F\'isica Aplicada, IGIC - Universitat Polit\`ecnica de Val\`encia, Gandia 46730 Spain}

\author[4]{I.~J.~Arnquist}
\affil[4]{Pacific Northwest National Laboratory, Richland, Washington 99354, USA}

\author[4]{D.~M.~Asner\thanks{now at Brookhaven National Laboratory}}

\author[5, 6]{D.~Baxter}
\affil[5]{Department of Physics and Astronomy, Northwestern University, Evanston, Illinois 60208, USA}
\affil[6]{Enrico Fermi Institute, KICP and Department of Physics, University of Chicago, Chicago, Illinois 60637, USA}

\author[7]{E.~Behnke}
\affil[7]{Department of Physics, Indiana University South Bend, South Bend, Indiana 46634, USA}

\author[8]{M.~Bressler}
\affil[8]{Department of Physics, Drexel University, Philadelphia, Pennsylvania 19104, USA}

\author[2]{B.~Broerman}

\author[2]{G.~Cao}

\author[5]{C.~J.~Chen}

\author[2]{U.~Chowdhury\thanks{now at Canadian Nuclear Laboratories}}

\author[2]{K.~Clark}

\author[6]{J.~I.~Collar}

\author[9]{P.~S.~Cooper}
\affil[9]{Fermi National Accelerator Laboratory, Batavia, Illinois 60510, USA}

\author[12]{C.~B.~Coutu}

\author[4]{C.~Cowles}

\author[4, 9]{M.~Crisler}

\author[2]{G.~Crowder}

\author[12]{N.~A.~Cruz-Venegas}
\affil[10]{Instituto de F\'isica, Universidad Nacional Aut\'onoma de M\'exico, M\'exico D.\:F. 01000, M\'exico}

\author[5, 9]{C.~E.~Dahl}

\author[11]{M.~Das}
\affil[11]{Astroparticle Physics and Cosmology Division, Saha Institute of Nuclear Physics, Kolkata, India}

\author[12]{S.~Fallows}
\affil[12]{Department of Physics, University of Alberta, Edmonton, T6G 2E1, Canada}

\author[13]{J.~Farine}
\affil[13]{Department of Physics, Laurentian University, Sudbury, P3E 2C6, Canada}

\author[3]{I.~Felis}

\author[14]{R.~Filgas}
\affil[14]{Institute of Experimental and Applied Physics, Czech Technical University in Prague, Prague, Cz-12800, Czech Republic}

\author[13, 15]{F.~Girard}
\affil[15]{D\'epartement de Physique, Universit\'e de Montr\'eal, Montr\'eal, H3C 3J7, Canada}

\author[2]{G.~Giroux}

\author[1]{J.~Hall}

\author[2]{C.~Hardy}

\author[16]{O.~Harris}
\affil[16]{Northeastern Illinois University, Chicago, Illinois 60625, USA}

\author[13]{T.~Hillier}

\author[4]{E.~W.~Hoppe}

\author[4]{C.~M.~Jackson}

\author[5]{M.~Jin}

\author[7]{L.~Klopfenstein}

\author[12]{C.~B.~Krauss}

\author[15]{M.~Laurin}

\author[1, 13]{I.~Lawson}

\author[13]{A.~Leblanc}

\author[7]{I.~Levine}

\author[13]{C.~Licciardi}

\author[9]{W.~H.~Lippincott}

\author[4]{B.~Loer}

\author[14]{F.~Mamedov}

\author[12]{P.~Mitra}

\author[2]{C.~Moore}

\author[7]{T.~Nania}

\author[8]{R.~Neilson}

\author[2]{A.~J.~Noble}

\author[7]{P.~Oedekerk}

\author[6]{A.~Ortega}

\author[12]{M.-C.~Piro}

\author[15]{A.~Plante}

\author[13]{R.~Podviyanuk}

\author[17]{S.~Priya}
\affil[17]{Materials Research Institute, Penn State, University Park, Pennsylvania 16802, USA}

\author[15]{A.~E.~Robinson}

\author[11]{S.~Sahoo}

\author[13]{O.~Scallon}

\author[11]{S.~Seth}

\author[9]{A.~Sonnenschein}

\author[15]{N.~Starinski}

\author[14]{I.~\v{S}tekl}

\author[2]{T.~Sullivan}

\author[15]{F.~Tardif}

\author[10, 13]{E.~V\'azquez-J\'auregui}

\author[7]{N.~Walkowski}

\author[13]{E.~Weima}

\author[13]{U.~Wichoski}

\author[4]{K.~Wierman}

\author[17]{Y.~Yan}

\author[15]{V.~Zacek}

\author[5]{J.~Zhang\thanks{now at Argonne National Laboratory}}

\date{November 2018}
\maketitle

\begin{abstract}
    The identification of non-signal events is a major hurdle to overcome for bubble chamber dark matter experiments such as PICO-60. The current practice of manually developing a discriminator function to eliminate background events is difficult when available calibration data is frequently impure and present only in small quantities. In this study, several different discriminator input/preprocessing formats and neural network architectures are applied to the task. First, they are optimized in a supervised learning context. Next, two novel semi-supervised learning algorithms are trained, and found to replicate the Acoustic Parameter (AP) discriminator previously used in PICO-60 with a mean of 97\% accuracy.
\end{abstract}

\pagebreak

\section*{Overview}

This paper is organized as follows:
\begin{itemize}
    \item In Section 1, we outline generally the PICO-60 experiment and the context in which machine learning is applicable.
    \item In Section 2, we discuss the specific properties of the experiment and its data, as well as the existing signal versus background discriminator.
    \item In Section 3, we discuss various machine learning techniques that are applied to develop a new discriminator.
    \item In Section 4, we document experimentation with supervised learning, applied to various data formats.
    \item In Section 5, we document experimentation with semi-supervised learning with the goal of improving data efficiency.
    \item In Section 6, we conclude this study and discuss its implications.
    \item In Section 7, we overview the technologies and software used for this study.
\end{itemize}

\section{Objective} \label{objective}

In the PICO-60 \cite{pico} experiment, and in other experiments striving to directly detect WIMP dark matter \cite{wimp}, one of the most important and challenging hurdles to overcome is that of background events. Unwanted particles from a variety of sources can produce event signatures that appear very similar to those which are expected to be created by dark matter candidates.

To resolve this, physicists must determine a discriminator function that can, based on experimental data, separate events generated by possible dark matter candidates from those due to background radiation. Two general problems stand in the way:

\begin{enumerate}
    \item A detailed model of the physical environment is often used to produce an accurate discriminator. This takes a long time to develop and compute, and will have to be updated whenever the physical variables of the experimental apparatus change.

    \item A major intent of many dark matter experiments is to reduce the number of background events to the lowest level possible. This means that there will be a very small number detected. This places a significant constraint on the amount of data that is available to optimize such a discriminator.

    Furthermore, what little data is available to optimize a discriminator very often contains impurities, because whatever background radiation is present during WIMP detection runs is also present during calibration runs. These are difficult to separate without already having access to a functional discriminator, creating a ``chicken or the egg'' problem (from which a common way to escape is to create a time- and labor-intensive simulation).
\end{enumerate}

The objective of this study is to investigate the potential for machine learning techniques to address both of these issues. More specifically, in the context of the PICO-60 experiment, there are two objectives:

\begin{enumerate}
    \item Determine the most effective machine learning architecture to use in development of a discriminator. Furthermore, develop several neural network architectures and apply them to different input/preprocessing formats (such as audio Fourier transforms, raw waveforms, and images), in order to find which are the most effective.
    \item Determine whether these techniques can be extended, applying two original algorithms based on semi-supervised learning, to develop an effective discriminator function based on incomplete and inaccurate initial information.
\end{enumerate}

For clarity, and with specific reference to the PICO-60 experiment, the intent is to accurately separate particle types using only calibration/background data collection run types (henceforth ``run types'') as training data. While run types are known to be mostly correlated with particle types, the relatively large percentage of impurities has the potential to hinder conventional learning techniques. Specifically, the run types available for training are:

\begin{enumerate}
    \item Neutron calibration sets, which are known to contain approximately 90\% neutrons, and
    \item Previously blinded WIMP search runs and diagnostic background runs, which contain approximately 99\% alpha particles.
\end{enumerate}

The architectures described in this paper may have wider applications to several similar cases, such as those that require a discriminator which is difficult to construct using conventional techniques. The following are possible reasons for this limitation:

\begin{enumerate}
    \item Limited ability to collect accurate calibration data, and/or imperfect correlation between calibration data and particle types.
    \item Unknown and/or complex interactions in the experimental apparatus make a conventional discriminator difficult to define, where machine learning techniques may be able to uncover hidden correlations in the input data.
\end{enumerate}

\section{Previous PICO-60 Analysis}

\subsection{Background}

\subsubsection{Introduction}

The PICO-60 \cite{pico} bubble chamber experiment, created for the detection of weakly interacting massive particles (WIMPs), was completed in 2017. It provides an excellent basis on which to develop and test the efficacy of new machine learning discriminators, relative to the techniques used during initial analysis of the experiment.

\subsubsection{About the Detector}

\begin{figure}[H]
    \centering
    \includegraphics[width=\textwidth]{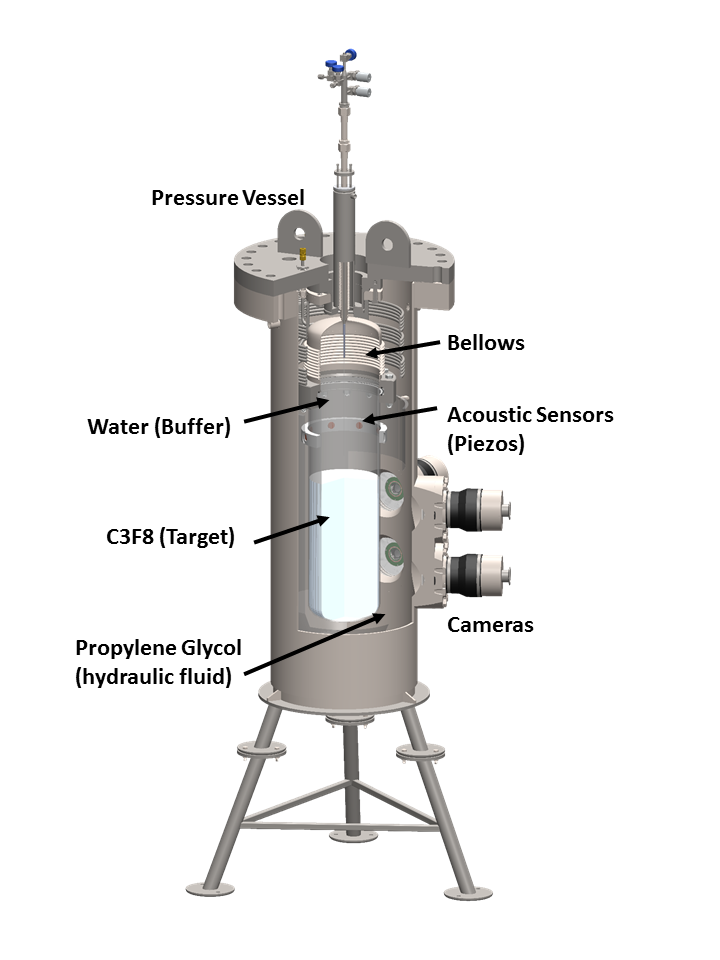}
    \caption{\label{pico_cad} CAD rendering of the PICO-60 detector as configured for its operation with C$_3$F$_8$.}
\end{figure}

The detector was filled with 52 kg of liquid C$_3$F$_8$, which was held at a constant temperature by a large chilled water bath. The pressure of the target liquid, held inside a synthetic quartz inner vessel, was precisely controlled with propylene glycol as hydraulic fluid inside a stainless steel pressure vessel (Figure \ref{pico_cad}).

The target was superheated by reducing its pressure until it entered a metastable state in which the vapor phase became energetically accessible. In the presence of a sufficiently dense energy deposition, a bubble would be nucleated and would grow to visible size. Four cameras stereoscopically monitored the chamber for these bubbles, issuing a recompression trigger upon their appearance. At this time, acoustic traces were recorded that captured the sound of bubble formation, and a Dytran 2006V1 fast pressure transducer \cite{amole_thesis} monitored the pressure rise. The rising pressure caused the vapor bubble to condense as the fluid reverted to a non-superheated state. Following this, the hydraulic system prepared to return the detector to the sensitive superheated state by once again expanding to a lower pressure.

Superheated fluid detectors have an intrinsic insensitivity to the relatively diffuse energy depositions from electron recoils \cite{pico}, and neutron backgrounds are mitigated by shielding the detector volume in a 20~t water tank. Alpha decays along the U/Th chains present the dominant remaining background which must be distinguished from nuclear recoils acoustically \cite{isotopes}.

\subsubsection{Properties of Events}

The nuclear recoils produced by WIMP candidates (in the cross-section and mass ranges to which the PICO-60 detector is sensitive) are predicted to be indistinguishable from those induced by neutrons. The key difference is that neutrons frequently (but not always) scatter several times and produce multiple bubbles, where the extremely small predicted cross-sections of WIMP candidates mean that they should almost invariably create just one bubble. This means single-bubble neutron events can be used to optimize a discriminator to detect WIMP events.

The approximate expected ratio between single-bubble and multiple-bubble events generated by neutrons is known based on Monte Carlo simulation and extrapolations from experience with previous detectors. A measured quantity of single-bubble events generated by nuclear recoils, in excess of this prediction, would be indicative of WIMP interactions. Consequently, it is imperative to isolate only those events associated with nuclear recoils.

\subsection{Discrimination Techniques}

\subsubsection{Overview}

In this study, all newly developed machine learning-based discriminators were compared to the two discrimination techniques used previously for the PICO-60 experiment: Acoustic Parameter and machine learning.

\subsubsection{Acoustic Parameter}

An Acoustic Parameter method (furthermore AP) is the technique that was used for event discrimination in the original PICO-60 analysis. A function was defined that discriminates between alpha particles and nuclear recoils based on audio data. It was derived based on a physical model of the bubble chamber, and was tuned using neutron calibration sources to produce events in the PICO-60 apparatus.

Before AP can be calculated, the audio data must first be converted into the frequency domain (specifically, with the banded Fourier transform $\beta_{8}$). It must also be preprocessed with the position correction function (discussed in Section \ref{camera_derived}), which adjusts the overall amplitude of an audio recording according to its position in the vessel. This is done because the acoustic characteristics differ significantly depending on the position of the event in the vessel \cite{pico}.

Finally, a set of data cuts must be applied to remove classes of events which are not accurately classified using AP alone. These classes include events originating close to the walls of the detector, and multiple-bubble events.

These conversion, preprocessing, and data cutting steps were reapplied during this study, and are described in Section \ref{data_formats}.

\subsubsection{Original Machine Learning Analysis} \label{og_analysis}

As part of the PICO-60 experiment, some testing was conducted to determine whether machine learning (in the form of a multi-layer perceptron) could be a viable technique to consider for a discrimination function. The results, published in the 2017 PICO-60 paper \cite{pico}, indicated that the general technique warranted further investigation.

The neural network in this study used the same $\beta_{8}$ input format as was applied in this study (detailed in Section \ref{piezo_derived}), with pre-trigger noise subtracted. Position corrections (also in Section \ref{piezo_derived}) were not applied; rather, the position of the event in the vessel is additionally used as input. AP was used to restrict the data used as input to the neural network. In addition to neutrons and alpha particles, gamma events were used as training and testing data for this neural network.

Note that it is possible there are position-related biases present in this training data, because neutron events during calibration runs occur more frequently near the wall of the detector where the calibration source is placed. These biases are expected to be alleviated by application of position corrections which normalize the audio amplitude; these corrections are detailed in Section \ref{piezo_derived}.

Software for this experiment was implemented using the MATLAB Neural Network Toolbox \cite{matlab}.

\subsection{Data Sources and Formats} \label{data_formats}

Discriminators can use a variety of different data formats to make predictions. These are collected from the two essential sensors present in the PICO-60 apparatus: two piezoelectric microphones (piezos) and cameras.

\subsubsection{Piezo-Derived} \label{piezo_derived}

\textit{Raw Waveform}

The lowest-level data derived from the two piezos in the PICO-60 apparatus is the raw waveform $\omega$. This consists of a series of 250,000 samples, collected at the rate of one per microsecond. Each sample is represented as a 16-bit integer. Only the section of the audio between sample 90,000 (inclusive) and sample 190,000 (exclusive) was used during most experiments, because there is only background noise prior to this window, and a clipped signal produced by hydraulics repressurizing the vessel after. These sections contain no information and can be undesirably fitted by neural networks. (For context, approximately half of the samples are recorded prior to the camera trigger.)

\textit{Fourier Transform}

The raw waveform $\omega$ was converted into the frequency domain by means of a one-dimensional Discrete Fourier Transform (DFT) for real input. Since DFTs produce sequences of complex numbers, the magnitude of each element was computed. The direct output from such a DFT is the full-resolution Fourier transform $\beta _{50,001}$, which consists of a sequence of 50,001 numbers (half the length of the raw waveform $\omega$).

Beyond this, the banded Fourier transform $\beta_{8}$ (which is the input to the AP function and the original machine learning analysis) was computed by integrating the resonant energy over all frequencies within each of a set of eight frequency bands ranging from 1 kHz to 300 kHz \cite{pico}. This can be thought of as a method of signal downscaling, compressing the information into a smaller number of data points.

Both $\beta _{50,001}$ and $\beta_{8}$ were used as inputs to machine learning models applied in this study.

\begin{figure}[h]
    \centering
    \includegraphics[width=\textwidth]{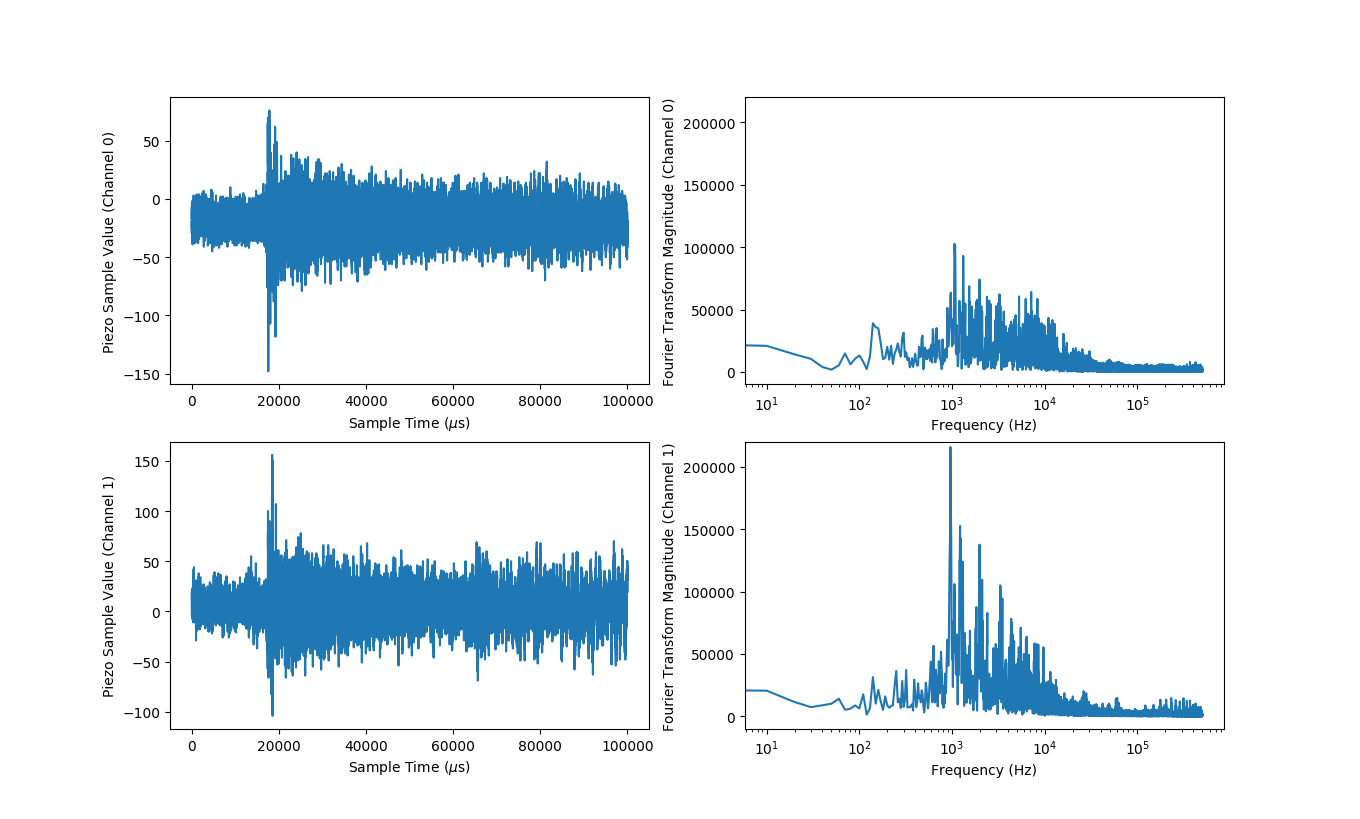}
    \caption{\label{} An example of the audio waveform $\omega$ with initial background and hydraulic noise cropped (left) and the resulting full resolution Fourier transform $\beta_{50,001}$ (right) for both audio channels. Recorded on 2016-12-03 at 04:22:45 UTC.}
\end{figure}

\subsubsection{Camera-Derived} \label{camera_derived}

\textit{Image Window Sequence}

Bubble events are detected using the four cameras within the PICO-60 apparatus, with an image-based trigger. When this happens, each camera records a sequence of 41 images before and after formation. These raw images contain a large amount of extraneous information; they encompass the entire vessel. To reduce the input information, the image window sequence $\iota$ includes 50\texttimes50 cropped windows around the position of the bubble. The 41 frames, many of which contain either no bubble or a bubble in later stages of formation, are reduced to 10 frames immediately around the formation of the bubble. Two to three frames from before the recording trigger are included in $\iota$, because the bubble does not cause a trigger until it is already of a significant size.

\begin{figure}[h]
    \centering
    \includegraphics[width=0.7\textwidth]{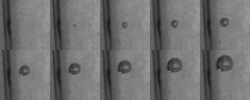}
    \caption{\label{} An example of the image sequence $\iota$. Recorded on 2016-09-23 at 22:15:44 UTC.}
\end{figure}

\textit{3D Position}

The 3-dimensional position $\chi$ of the bubble within the vessel is calculated using triangulation, based on the known positions and angles of the cameras and the position of the bubble within the field of view of each camera.

This is used in the banded frequency position correction function $PosCor(\beta _{8}, \chi)$, which corrects $\beta _{8}$ for variations in amplitude which depend on the position of the bubble. $PosCor(\beta _{8}, \chi)$ is used for training several machine learning models throughout this study. Details of its use are in Section \ref{mlp}.

\subsection{Data Cuts}

Techniques in this study are trained and validated on a number of different data sets. The selection of these sets focused on the fundamental trade-off between quality and quantity of data; by setting a higher standard for the validity of training data, one has less data to train on.

The initial data set $D$ to which these cuts are applied consists of all events recorded in the PICO-60 detector, at the 3.3keV threshold, between 2016-09-02 and 2017-06-20.

\subsubsection{Basic Quality Cut}

A number of cuts are necessarily applied to all data to ensure meaningful results. Otherwise, significant overfitting on biases in the data is likely. This basic quality cut {\it QualCut}$(D)$ consists of the following restrictions:

\begin{itemize}
    \item The run was not collected during engineering or testing
    \item The recording process was triggered by the camera (as opposed to a manual trigger, et cetera)
    \item AP is not erroneously large and negative ($log_{10}(AP)>-100$)
    \item The event was recorded more than 25 seconds after reaching target pressure
    \item The bubble position $\chi$ was successfully calculated ($[\chi_{X}, \chi_{Y}, \chi_{Z}]\neq[-100, -100, -100]$)
\end{itemize}

\subsubsection{Bubble Multiplicity Cut}

PICO-60 events which include multiple bubbles are \textit{always} neutron events; alpha particles are stopped electromagnetically, and WIMP candidates have a negligible probability of multiple scattering because of their extremely low predicted cross section. Thus, multi-bubble events can be safely removed. The bubble multiplicity cut {\it MultiCut}$(D)$ consists of the following restrictions:

\begin{itemize}
    \item Either 0 or 1 bubbles are detected based on images from the camera
    \item The number of bubbles approximated using the fast pressure transducer is close to 1
\end{itemize}

\subsubsection{Wall Cut} \label{wall_cut}

Events that occur near the walls of the vessel have acoustic properties which are very different from events nearer the center of the vessel. It can be desirable for a discriminator to handle these events correctly; however, AP alone does not, and neither does the neural network used in the previous PICO-60 paper. Thus, removing wall events allowed for a more meaningful direct comparison between a new neural network and existing techniques.

\begin{figure}[h!]
    \centering
    \includegraphics[width=\textwidth]{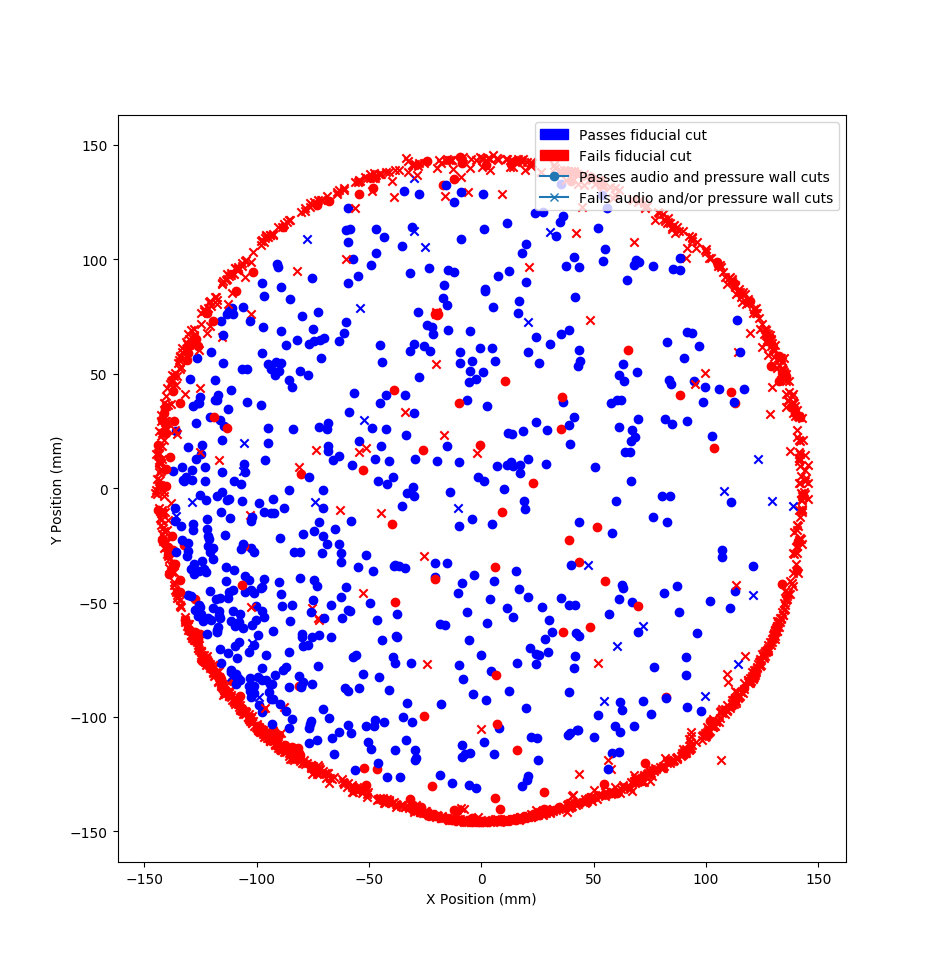}
    \caption{\label{wall_cuts_visualization} A visualization of the fiducial cuts and how they match up with the pressure and acoustic cuts.}
\end{figure}

The complete wall cut {\it WallCut}$(D)$ is a composition of the fiducial cut {\it FidWallCut}$(D)$ (which makes use of the 3D position $\chi$), the pressure cut {\it PresWallCut}$(D)$ (which uses data from the fast pressure transducer), and the acoustic cut {\it AcWallCut}$(D)$ (which uses the banded Fourier transform $\beta _{8}$). Those cuts restrict data as follows:

\begin{itemize}
    \item The fiducial cut {\it FidWallCut}$(D)$ defines a spatial area along the walls of the vessel within which no events are accepted. It determines whether events fall within this area based on the visually calculated position $\chi$.
    \item The pressure cut {\it PresWallCut}$(D)$ restricts the pressure detected by the fast pressure transducer, without any position corrections, to be within a range of 0.3 of 1
    \item The acoustic cut {\it AcWallCut}$(D)$ is defined using the banded Fourier transform $\beta_{8}$, and takes advantage of differences in the frequency distribution (specifically the first and second frequency bands of the first and third piezos) of wall events and non-wall events.
\end{itemize}

In Figure \ref{wall_cuts_visualization}, the relative and combined effectiveness of each of these cuts can be seen. It is clear that most of the time, both of these cuts agree, but there are certain cases in which only one or the other eliminates an event.

Some events eliminated by fiducial cuts appear to be in the middle of the vessel in Figure \ref{wall_cuts_visualization}. These events are at the top or bottom of the detector (along the Z axis, which is not displayed in the 2D plot).

The data set, after all of these cuts are applied, consists of 624 events. During each training run, 128 randomly selected events are set aside for validation, leaving a training set of 496 events.

\section{Machine Learning Techniques}

\subsection{Background}

While neural networks have existed for many years, recent developments in computing power, most importantly GPU acceleration \cite{gpu}, have opened up a wide variety of applications and fields in which they can now be used.

One of the key benefits of machine learning systems is that they have the capacity to approximate arbitrary functions (such as AP) without human intervention. This greatly alleviates the need for human programmers to define and optimize the exact algorithm used.

This characteristic has already shown wide-ranging implications in many fields, such as autonomous driving \cite{end-to-end} and detection of heart arrhythmia \cite{arrhythmia}. For one, it has the potential to drastically increase the speed of iteration for the people working on a project, since retraining a machine learning system when a hardware or software variable changes is much faster than calibrating a human-designed model.

\subsection{Techniques Considered}

Machine learning techniques are usually divided into two main categories:

\begin{itemize}
    \item Supervised learning, in which a system is trained on a set of fully-labeled data to classify unseen examples according to the patterns it observes in the training set, and
    \item Unsupervised learning, in which a system finds patterns or clusters in a set of unlabeled data. This can find order in almost any training set, but it is unpredictable and may not find the particular patterns desired.
\end{itemize}

A significant challenge in this application is that neither supervised nor unsupervised learning is ideal. The calibration and background runs available for training are not pure, so supervised learning is likely to overfit on biases and produce undesirable results. Conversely, unsupervised learning may be able to find clusters in the data, but it is impossible to guarantee that it will distinguish between nuclear recoils and alpha particles as opposed to some other binary separation.

For these reasons, a technique is needed that is not as sensitive to problematic training data as supervised learning, but more predictable and controllable than unsupervised learning. Semi-supervised learning is a middle ground in this regard. It makes use of a labeled set as well as an unlabeled set. It uses the labeled set for training, and uses the unlabeled set to further structure the patterns it finds in the training set (which may not be sufficiently large to apply to supervised learning).

Two original semi-supervised algorithms were developed and implemented for this study: gravitational differentiation (GD) and iterative cluster nucleation (ICN). Both are discussed in detail in Section \ref{semi_supervised}.

\subsection{Performance Analysis}

Performance of the machine learning systems was evaluated using two metrics: classification accuracy and class-wise standard deviation.

Classification accuracy is the ability of the network to separate the events into the two desired classes. Numerically, this corresponds to the number of validation examples the network classifies correctly. The baseline for correctness is the run type during supervised learning experiments, and the AP prediction for semi-supervised experiments.

The correctness of the network's predictions is further detailed in the precision and recall statistics. Precision represents the proportion, of those that the network predicts to be nuclear recoils, that are actually nuclear recoils (or part of the corresponding calibration set). It is an indicator of how pure the network's predictions are for the recoil class. Recall represents the proportion, of those that are actually nuclear recoils, that the network predicts to be nuclear recoils. It indicates how many examples the network misses (i.e., those that would erroneously be removed as background radiation).

Class-wise standard deviation, abbreviated CWSD, captures how decisive the network is: whether its prediction is confidently high or low, or is only slightly closer to one edge of the spectrum.

Numerically, this is a variable defined below, where $N$ and $A$ are the sets of outputs of the supervised learning discriminator in question, corresponding to the sets of neutrons and alpha particles respectively (or calibration and background sets respectively, depending on which is used as ground truth data). It calculates the spread of the network's predictions for each ground truth class. This means that a decisive discriminator, which produces a wide separation between the two classes, is preferred over one that produces a nebulous cloud of outputs with a seemingly arbitrary decision boundary.

\begin{equation}
    S=std(N \cup A)
\end{equation}
\begin{equation}
    \mbox{\it CWSD}=(std(N \div S) + std(A \div S)) \div 2
\end{equation}

The first step is to calculate the standard deviation $S$ of the union of $N$ and $A$. This gives an indication of the scale of the overall distribution. When $N$ and $A$ are divided by $S$, they are normalized so that the standard deviation of their union is equal to 1. While the neural network's outputs are bounded in the range of 0 to 1 with a sigmoid activation on the last layer, AP has a significantly wider range. Normalization of the union prevents this from creating a bias where AP would produce a higher standard deviation with a similarly proportioned error.

The second step is to calculate the mean of the standard deviations of the normalized sets of neutrons and alpha particles individually. This is an indication of how tightly clustered or widely dispersed the discriminator's predictions are for each class. Very consistent predictions of $x$ for neutrons and $y$ for alphas (for any given $x$ and $y$), with minimal variance off those specific values, will produce a low class-wise standard deviation.

\subsection{Optimization Process}

Each of the learning algorithms used in this study applies a neural network. For each of the general configurations (multi-layer perceptron, one-dimensional convolutional neural network, et cetera) there are many possible specific architectures with different hyperparameters, which have to be optimized to improve performance. Specific hyperparameters that were optimized include:

\begin{itemize}
    \item Number of layers of each type (dense, convolutional)
    \item Number of neurons in dense layers
    \item Number of filters in convolutional layers (depth of output tensor)
    \item Kernel size in convolutional layers (spatial area that a single filter covers)
    \item Stride in convolutional layers (spacing between kernel positions during convolution)
    \item Dropout regularization parameter (proportion of neurons to randomly remove for any given training example)
    \item L2 regularization $\lambda$ (multiplier for squared weights before adding to loss function)
\end{itemize}

Several of these hyperparameters are optimized at a time, using a grid search. Given $n$ different hyperparameters to optimize and $m$ different possible values for each, $m^n$ different networks are trained (using every possible combination of hyperparameters) and tested on a validation set. For space reasons, only a small subset of the tested configurations are displayed in tables throughout this paper.

In general, the parameters for each grid search were chosen based on some initial empirical experimentation to determine the general range within the entire parameter space that work reasonably well. There was a limit to how many parameters could be tested, especially in the case of semi-supervised learning, as some of the grid searches took several days to run.

Where not otherwise specified, the following hyperparameters were used universally:
\begin{itemize}
    \item Adam \cite{adam} stochastic optimizer
    \item Mean squared error loss function
    \item Batch size of 32
    \item $tanh$ activation function
\end{itemize}

\section{Supervised Learning}

\subsection{Overview}

The primary objective, in experimentation with supervised learning, is to determine the most effective combination of input format and neural network architecture to use in replicating a discriminator function. Three major configurations were tested:

\begin{enumerate}
    \item A convolutional neural network trained on the raw waveform $\omega$.
    \item A dense neural network trained on banded Fourier transforms $\beta_{N}$.
    \item A convolutional neural network trained on the image window data $\iota$.
\end{enumerate}

All supervised learning configurations were trained and evaluated based on their ability to determine the origin of events: whether they are from neutron calibration source runs (predominantly neutrons) or background radiation runs (predominantly alpha particles). This metric was used for training and testing instead of AP for two reasons:

\begin{enumerate}
    \item During early operations of future experiments, when no AP equivalent is yet available, impure data is the only information available for training. The network structure should be selected to accommodate this.
    \item In general, supervised learning systems learn to replicate biases present in the training data set. Thus, to evaluate the network's effectiveness at processing the input data, the run type (which is used for training) must be used as a baseline instead of AP (which is never provided to the network during the training process).
\end{enumerate}

\subsection{Convolutional Neural Network for Raw Waveform Analysis}

\subsubsection{Structure}

A convolutional neural network was used for analyzing the raw waveform $\omega$ directly, without any preprocessing whatsoever. It is furthermore referred to as $DeepConv(\omega)$. Direct processing of the waveform avoids any destruction of information; theoretically, a sufficiently complex neural network should be capable of deriving any characteristics needed.

For this task, a very deep (20 layers in its shallowest configuration) one-dimensional fully convolutional neural network was applied. The architecture was inspired by the M34-res network \cite{verydeepconvnets} for analysis of raw waveforms. L2 regularization was used to alleviate overfitting, and batch normalization was applied to the input to stabilize training.

The following hyperparameters were tested during a grid search:
\begin{itemize}
    \item L2 Regularization $\lambda \in \{0.0003, 0.001, 0.003\}$
    \item Dense Layer Dropout $\in \{0, 0.25, 0.5\}$
    \item Convolutional Filters $\in \{24, 48\}$
    \item Convolutional Kernel Size $\in \{3, 5\}$
    \item Number of Convolutional Layers $\in \{17, 32\}$
    \item Number of Dense Layers = 3 (with neuron counts 64, 16, and 1)
\end{itemize}

\subsubsection{Results}

\begin{figure}[h]
    \centering
    \includegraphics[width=\textwidth]{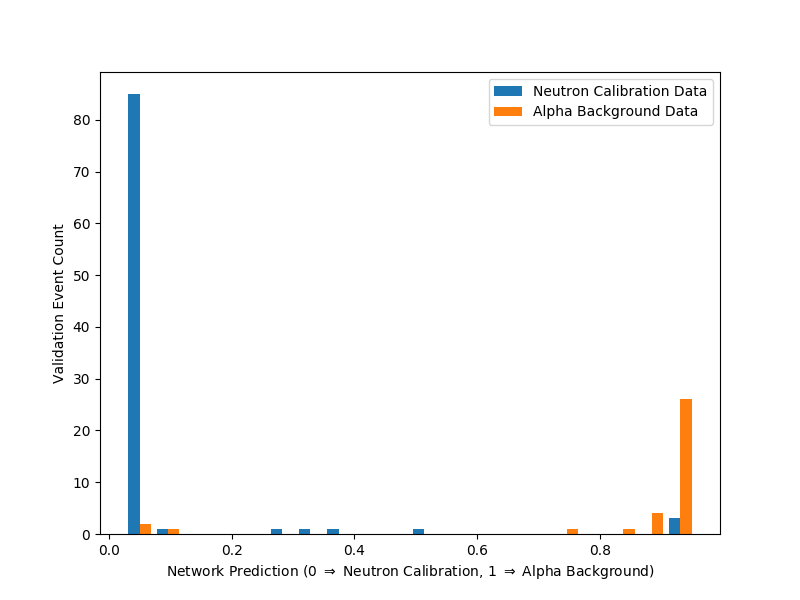}
    \caption{\label{waveform_hist} Prediction distribution of the best {\it DeepConv}$(\omega)$ discriminator. (Validation Event Count is the number of events in the validation set that fall within a certain network prediction band.)}
\end{figure}

The most accurate configuration during a waveform CNN grid search was 95\% accurate and had a standard deviation of 0.52 (seen in Figure \ref{waveform_hist}).

Table \ref{deepconv_table} shows the results of several {\it DeepConv}$(\omega)$ discriminators produced throughout testing, with corresponding network hyperparameters.

\begin{minipage}{\textwidth}
    \begin{center}
        \captionof{table}{Performance of {\it DeepConv}$(\omega)$ configurations.} \label{deepconv_table} 
        \begin{tabular}{|l|l|l|l|l|l|l|l|l|}
            \hline
            L2 $\lambda$ & Dropout & Filters & Kernel & Conv Layers & Max Accuracy & Precision & Recall & CWSD \\
            \hline
            0.003 & 0 & 24 & 3 & 17 & 95\% & 97\% & 97\% & 0.52 \\
            \hline
            0.003 & 0.25 & 48 & 3 & 17 & 94\% & 100\% & 91\% & 0.44 \\
            \hline
            0.001 & 0.25 & 24 & 5 & 17 & 95\% & 99\% & 94\% & 0.46 \\
            \hline
            0.001 & 0.25 & 24 & 5 & 32 & 91\% & 94\% & 94\% & 0.66 \\
            \hline
            0.0003 & 0 & 48 & 5 & 32 & 88\% & 94\% & 89\% & 0.61 \\
            \hline
            0.0003 & 0 & 24 & 3 & 17 & 93\% & 96\% & 95\% & 0.59 \\
            \hline
        \end{tabular}
    \end{center}
\end{minipage}

Table \ref{deepconv_table} shows that simple network architectures seem to produce better accuracy, likely due to reduced overfitting. Larger kernel sizes, more filters, and more layers make performance generally worse, while higher L2 regularization does the opposite. Dropout seems to have a weak correlation, possibly reducing accuracy slightly.

\subsection{Multi-Layer Perceptron for Fourier Transform Analysis} \label{mlp}

\subsubsection{Structure}

Preprocessing the audio by applying a Fourier transform may be advantageous. If the frequency distribution and overall amplitude are indeed the most important factors for discrimination, the network can gather them straight from the Fourier transform $\beta_{N}$ rather than having to analyze $\omega$ to extract this information.

In addition to the banded Fourier transform $\beta_{8}$ used in the original PICO-60 analysis, the full-resolution $\beta _{50,001}$ was input into a neural network directly, without any downscaling.

When $\beta_{8}$ was used as input, position corrections were optionally applied to the input data, as they were the input to Acoustic Parameter. The resulting configurations are furthermore referred to as $FourierMLP(\beta_{N})$ and $FourierMLP(PosCor(\beta_{N}, \chi))$. While a variety of network architectures were tested, most had a small number of dense layers (on the order of three), applied dropout and L2 regularization, and used batch normalization once again.

The following hyperparameters were tested during a grid search:
\begin{itemize}
    \item Dropout $\in \{0, 0.25, 0.5\}$
    \item L2 Regularization $\lambda \in \{0, 0.0003, 0.001, 0.003, 0.01\}$
    \item Number of Dense Layers $\in \{2, 3, 4\}$
\end{itemize}

\subsubsection{Results}

\begin{figure}[h]
    \centering
    \includegraphics[width=\textwidth]{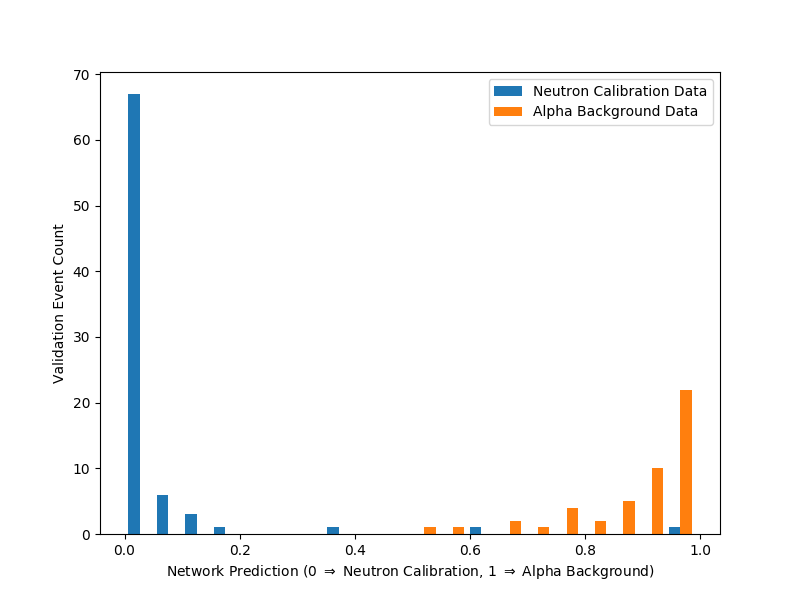}
    \caption{\label{banded_no_pos_input_hist} Prediction distribution of the best {\it FourierMLP}$(\beta_{8})$ discriminator.}
\end{figure}

It became evident very quickly that a high resolution was not required to get a high accuracy relative to the ground truth data. {\it FourierMLP}$(\beta_{8})$ managed an excellent 98\% validation accuracy, and also produced a mean class-wise standard deviation of 0.29. Its very decisive prediction distribution can be observed in Figure \ref{banded_no_pos_input_hist}.

Very interestingly, {\it FourierMLP}$(\beta_{8})$ performed better than {\it FourierMLP}$(${\it PosCor}$(\beta_{8}), \chi))$, which produced an accuracy value of 96\%. Both of these results were better than any of the network architectures trained on $\omega$.

Table \ref{fft_wall_cuts} compares various multi-layer perceptron discriminators, with and without wall cuts.

\begin{minipage}{\textwidth}
    \begin{center}
        \captionof{table}{Performance of {\it FourierMLP}$(\beta_{N})$ configurations.} \label{fft_wall_cuts}
        \begin{tabular}{|l|l|l|l|l|l|l|l|}
            \hline
            Configuration & L2 $\lambda$ & Dropout & Layers & Max Acc & Prec & Recall & CWSD \\
            \hline
            {\it FourierMLP}$(${\it PosCor}$(\beta_{8}, \chi))$ & 0 & 0.5 & 2 & 96\% & 98\% & 97\% & 0.42 \\
            \hline
            {\it FourierMLP}$(\beta_{8})$ & 0 & 0.5 & 2 & 98\% & 100\% & 98\% & 0.29 \\
            \hline
            {\it FourierMLP}$(\beta_{50,001})$ & 0 & 0 & 2 & 95\% & 94\% & 98\% & 0.47 \\
            \hline
            {\it FourierMLP}$(\beta_{50,001})$ & 0.001 & 0 & 4 & 95\% & 95\% & 98\% & 0.46 \\
            \hline
            {\it FourierMLP}$(\beta_{50,001})$ & 0.001 & 0.5 & 4 & 94\% & 93\% & 98\% & 0.54 \\
            \hline
            {\it FourierMLP}$(\beta_{50,001})$ & 0.003 & 0.25 & 3 & 93\% & 92\% & 98\% & 0.57 \\
            \hline
        \end{tabular}
    \end{center}
\end{minipage}

It is very clear that higher resolution data does not improve accuracy in a relatively shallow network. Regularization, both dropout and L2, appears to have relatively little effect.

\subsection{Convolutional Neural Network for Image Window Analysis}

\subsubsection{Structure}

It is an open question whether or not there is any information in the image data $\iota$ that could be used to distinguish between particle classes. Relative to the extremely short period of time in which the bubble forms (on the scale of nanoseconds), the framerate of the camera (340Hz) is extremely slow. While the very early stages of bubble formation (when the sound is produced) are known to differ depending on whether the bubble was created by a nuclear recoil or an alpha particle, it is not experimentally known whether any visually apparent differences persist when the bubble is visible.

In an effort to resolve this, a two-dimensional convolutional neural network was applied to the task of discriminating based on $\iota$, referred to as {\it ImageConv}$(\iota)$. The network architecture consists of a moderate number of convolutional layers (on the scale of nine) and three dense layers at the end. L2 regularization was used on all layers, and dropout was additionally used on the dense layers.

The following hyperparameters were tested during a grid search:
\begin{itemize}
    \item Dropout $\in \{0, 0.25, 0.5\}$
    \item L2 Regularization $\lambda \in \{0, 0.0003, 0.001, 0.003, 0.006, 0.01\}$
    \item Number of Convolutional Layers $\in \{6, 9, 12\}$
    \item Number of Dense Layers = 3 (with neuron counts 64, 16, and 1)
\end{itemize}

\subsubsection{Results}

\begin{figure}[h]
    \centering
    \includegraphics[width=\textwidth]{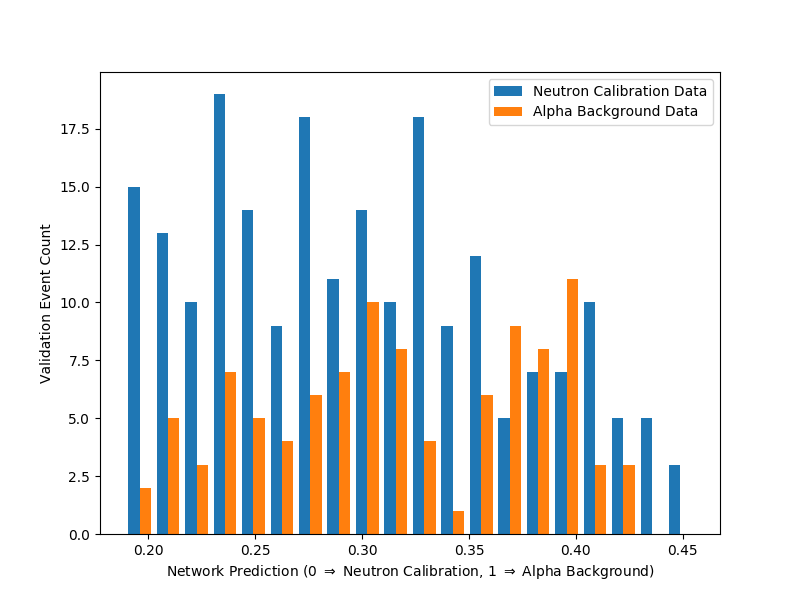}
    \caption{\label{image_hist} Prediction distribution of one of the best {\it ImageConv}$(\iota)$ discriminators. (Note the truncated X axis.)}
\end{figure}

Throughout a variety of different network architectures, the best validation accuracy obtained was 69\%. The corresponding training accuracy of 100\% indicates that severe overfitting is taking place (not unexpected, given the relatively small image data set). Also, the class-wise standard deviation was a very high 0.99, which implies its decisions are nearly random (as seen in Figure \ref{image_hist}). It is likely fitting poorly on some form of noise within $\iota$. The fact that throughout many trials during a grid search, no good performance on validation data was observed, provides significant evidence that images provide insufficient information for effective discrimination.

\begin{minipage}{\textwidth}
    \begin{center}
        \captionof{table}{Performance of {\it ImageConv}$(\iota)$ configurations.} \label{imageconv_table}
        \begin{tabular}{|l|l|l|l|l|l|l|l|}
            \hline
            L2 $\lambda$ & Dropout & Convolutional Layers & Dense Layers & Max Accuracy & Precision & Recall & CWSD \\
            \hline
            0 & 0 & 6 & 3 & 69\% & 69\% & 100\% & 0.99 \\
            \hline
            0.01 & 0.25 & 6 & 3 & 68\% & 68\% & 100\% & 0.98 \\
            \hline
            0.003 & 0.25 & 9 & 3 & 68\% & 68\% & 100\% & 0.98 \\
            \hline
            0.01 & 0.25 & 12 & 3 & 68\% & 68\% & 100\% & 0.97 \\
            \hline
            0.01 & 0.5 & 12 & 3 & 68\% & 68\% & 100\% & 0.99 \\
            \hline
        \end{tabular}
    \end{center}
\end{minipage}

In Table \ref{imageconv_table}, high standard deviations across the board are further evidence for the hypothesis that there is insufficient information in the image data. The 100\% recall statistics are not due to high classification accuracy, but rather due to the network making predictions less than 0.5 for all examples of both classes.

\subsection{Performance Summary}

Table \ref{super_overview} compares the accuracy and mean class-wise standard deviation values of certain successful supervised learning configurations with multiple input formats.

\begin{minipage}{\textwidth}
    \begin{center}
        \captionof{table}{Summary of supervised learning configurations.} \label{super_overview}
        \begin{tabular}{|l|l|l|l|l|}
            \hline
            Configuration & Max Accuracy & Precision & Recall & CWSD \\
            \hline
            {\it DeepConv}$(\omega)$ & 95\% & 97\% & 97\% & 0.52 \\
            \hline
            {\it FourierMLP}$(${\it PosCor}$(\beta_{8}))$ & 96\% & 98\% & 97\% & 0.42 \\
            \hline
            {\it FourierMLP}$(\beta_{8})$ & 98\% & 100\% & 98\% & 0.29 \\
            \hline
            {\it FourierMLP}$(\beta_{50,001})$ & 95\% & 95\% & 98\% & 0.46 \\
            \hline
            {\it ImageConv}$(\iota)$ & 68\% & 68\% & 100\% & 0.97 \\
            \hline
        \end{tabular}
    \end{center}
\end{minipage}

The key conclusion to be drawn from the above table is that the low-resolution Fourier transform $\beta_{8}$ is by far the most effective input format. Higher resolutions seem to make performance worse, likely due to overfitting.

Based on this result, experimentation with semi-supervised learning proceeded using \\ {\it FourierMLP}$(${\it PosCor}$(\beta_{8}, \chi))$, applying a three-layer perceptron architecture. There were three main reasons for this:

\begin{enumerate}
    \item This was a highly effective configuration, producing 96\% accuracy and a class-wise standard deviation dramatically better than the original PICO-60 neural network on that data set.
    \item The use of the exact same input format as AP makes the algorithms more directly comparable.
    \item If position correction is not applied, there is some risk of introducing a bias to the network's training data because there are different position distributions of the two classes (as described in Section \ref{og_analysis}).
\end{enumerate}

\section{Semi-Supervised Learning} \label{semi_supervised}

\subsection{Overview}

Semi-supervised learning is an uncommon machine learning technique, relative to widely-used supervised and unsupervised learning. The concept is to train a machine learning model on a set of labeled (classified) data in addition to a set of unlabeled (unclassified) data.

In the context of PICO-60, the labeled sets consist of the neutron calibration runs, and the background radiation runs (which consist predominantly of alpha particles). In general, these labeled sets should have a strong but not perfect correlation with particle types. The unlabeled data can consist of any mixture of particle types.

The primary advantage of semi-supervised learning is clear: it requires fewer labels to be collected than conventional supervised learning. A less obvious but powerful secondary advantage is that, since the learning process allows the network to reinforce its own decisions, the negative effect of impure training data is minimized, potentially allowing the resulting network to perform better than a network trained with supervised learning.

Two techniques for semi-supervised learning have been developed for this study: gravitational differentiation and iterative cluster nucleation. The essence of both algorithms is a positive feedback loop. First, the network is trained on a smaller amount of imperfect data. Throughout the training process, the network runs predictions on unlabeled data. By using its most confident predictions in the training process, the desired function (separating particle types) is continuously reinforced, while less confident predictions are de-emphasized. This effectively draws new, unlabeled data into the training set.

Because all data in the PICO-60 experiment is labeled, unlabeled training data sets are randomly sampled from the 496 training examples remaining (after the validation set is selected).

Since the intent of these techniques is to separate particle types (the same as AP), performance evaluations (accuracy and class-wise standard deviation) use AP as a baseline. All graphs in this section use AP's predictions to define the sets of neutrons and alpha particles.

To ensure that accuracy statistics are as consistent and reliable as possible, every configuration was trained three times, with a different randomly selected validation set each time. (Using more than three training runs would have taken prohibitive amounts of time.) Accuracy, precision, recall, and class-wise standard deviation statistics are the mean over those three tests.

\subsection{Gravitational Differentiation}

\subsubsection{Concept}

Gravitational differentiation (GD) is a novel technique, developed during this study, for training a neural network in a semi-supervised fashion on a relatively small set of imperfectly labeled data, while simultaneously incorporating the network's changing predictions on a set of unlabeled data to encourage decisive classifications.

In general terms, it amplifies the training effect associated with high-confidence predictions (those that most clearly show characteristics of either an alpha particle or a neutron) on unlabeled examples. Meanwhile, it minimizes the training impact of low-confidence predictions, such that the network will not be trained to make incorrect predictions.

Analogous to a gravitational effect, events with high-confidence predictions are attracted closer to their corresponding classifications. They are then used as training data, allowing the patterns they represent to be used more generally to make predictions on other events.

The system accomplishes this using a new method for calculating gradients for the last layer of the neural network. Based on the current predictions of a trained network on a certain unlabeled training example, a gradient is calculated for the last layer with the following purposes:

\begin{itemize}
    \item For examples with confident predictions close to 0 or 1: Create a large gradient that pulls the prediction closer to 0 or 1 through gradient descent (like gravity).
    \item For examples with low-confidence predictions close to 0.5: Create a gradient near 0 that does not affect training significantly.
\end{itemize}

The gradients produced by this technique are used for training in the same way as gradients calculated with a loss function. Backpropagation is used to calculate gradients for previous layers, and a stochastic gradient descent optimizer \cite{sgd} is used for weight updates.

\subsubsection{Algorithm}

Gradients are calculated using the gravitational differentiation function $\mathrm{GravDiff}(p, \psi, g)$, which is parameterized by the network's existing prediction $p$ on the training example in question, the degree $\psi$ of the piecewise exponential function used to distort the response of the gradient, and the gravitational multiplier $g$. The function is defined as follows:

\begin{equation}
    \mathrm{GravDiff}(p, \psi, g) = g \cdot \mathrm{sgn}(p) \cdot \mathrm{abs}(\mathrm{tanh}(2(p - 0.5))) ^ \psi
\end{equation}

It is a distortion of the hyperbolic tangent that flattens the central range and comparatively exaggerates the asymptote on either side. The equation above can be described in the following steps:

\begin{enumerate}
    \item Transform the network's prediction $p$, so its range is 1 to -1 rather than 0 to 1.
    \item Apply the sigmoidal hyperbolic tangent, producing a value of 0 in the center and asymptotic slopes to -1 and 1 at the edges.
    \item \label{exponent} Apply a relatively large exponent which is $\psi$ (in the range of 3 to 11) to squash values close to 0 closer to 0. This ensures a shallow slope in the center (so low-confidence network outputs near 0.5 will produce very small output gradients).
\end{enumerate}

(Note that step \ref{exponent} only works if $\psi$ is an odd integer (because odd integer exponents preserve the sign), limiting the ability to fine-tune this function by changing $\psi$. To resolve this, a piecewise exponential function is used instead, where the absolute value is taken prior to applying the power and the sign is multiplied back in afterwards. This permits use of the full range of curves produced by even and non-integer values of $\psi$.)

\subsubsection{Application}

This function is applied in a training algorithm that is parameterized by the set of imperfectly labeled data $\varsigma$, the set of unlabeled training data $\upsilon$, a binary classification neural network $NN(x)$ (where $x$ is the input data format), and the gravitational multiplier increment $\delta_g$. The distortion power $\psi$ is assumed to be constant throughout a training run, and the gravitational multiplier $g$ is initialized at 0, gradually increasing during training. It iterates as follows:

\begin{enumerate}
    \item Train $NN(x)$ for a single epoch on the combined set $\varsigma \cup \upsilon$, using the predefined ground truths for $\varsigma$ (with a mean squared error loss function) and the most recently calculated gravitational gradients for each of the examples in $\upsilon$.
    \item Calculate predictions $p$ with $NN(x)$ on the entirety of $\upsilon$. Calculate $\mathrm{GravDiff}(p, \psi, g)$ and record the resulting gradients for the next iteration.
    \item Increment $g$ by $\delta_{g}$. At the beginning, when $g = 0$, training will be entirely based on the labeled set $\varsigma$, and progressively, over the course of a run, the gravitational effect increases with $g$.
\end{enumerate}

This system was trained using a multi-layer perceptron with an 8-band Fourier transform input for $NN(x)$, because this was one of the most successful configurations in supervised learning. It produced the highest accuracy as well as the lowest class-wise standard deviation.

\begin{figure}[H]
    \centering
    \includegraphics[width=0.8\textwidth]{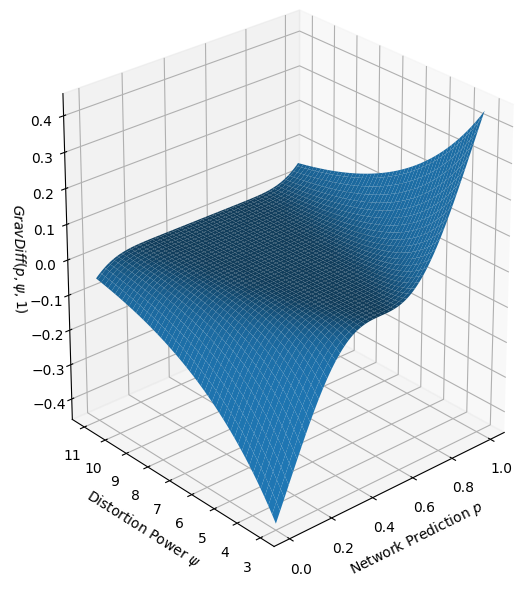}
    \caption{\label{grav_diff} A visualization of $\mathrm{GravDiff}(p, \psi, g)$ with respect to $p$ and $\psi$.}
\end{figure}

\subsubsection{Results}

During a grid search of parameters of the gravitational differentiation algorithm, in addition to the stochastic gradient descent learning rate (a multiplier that defines how quickly and precisely optimization occurs), the highest mean accuracy over three runs of an individual configuration was 99\%. The highest accuracy was obtained with 256 labeled examples in the set $\varsigma$ (out of 496 total training examples), which equates to 52\% of the data being labeled. High accuracy statistics in this range were also obtained using 128 examples (99\% accuracy, once again) and 64 examples (97\% accuracy). 32 examples produced a much lower maximum accuracy, at only 83\%. From this, it is evident that it is possible to achieve very high accuracy using this technique. It is also clear that, in general, larger initial sets produce higher performance.

However, the four hyperparameters optimized during the grid search were observed to have a weak polynomial correlation with the network's accuracy. This implies that the training process is generally insensitive to variations in those particular network parameters.

This lack of correlation also means that the mean accuracy over all configurations with a given $\varsigma$ size is a meaningful statistic, since the validation accuracy for an individual configuration is difficult to confidently predict. These statistics, shown in Table \ref{grav_accuracy_by_varsigma}, show that for all four set sizes, there is a fairly wide spread between maximum and mean accuracy.

\begin{minipage}{\textwidth}
    \begin{center}
        \captionof{table}{Gravitational differentiation accuracy statistics by size of $\varsigma$.} \label{grav_accuracy_by_varsigma}
        \begin{tabular}{|l|l|l|l|}
            \hline
            Initial $\varsigma$ Size & Proportion of Data Labeled & Max Accuracy & Mean Accuracy \\
            \hline
            32 & 6\% & 83\% & 69\% \\
            \hline
            64 & 13\% & 97\% & 73\% \\
            \hline
            128 & 26\% & 99\% & 84\% \\
            \hline
            256 & 52\% & 99\% & 94\% \\
            \hline
        \end{tabular}
    \end{center}
\end{minipage}

Table \ref{grav_overview} represents the performance of several high-accuracy gravitational differentiation configurations tested during the grid search. The accuracy statistic is the mean over three tests of the same configuration.

\begin{minipage}{\textwidth}
    \begin{center}
        \captionof{table}{Performance of gravitational differentiation configurations.} \label{grav_overview}
        \begin{tabular}{|l|l|l|l|l|l|l|l|}
            \hline
            $\varsigma$ Size & $\delta g$ & Learning Rate & $\psi$ & Mean AP Accuracy & Precision & Recall & CWSD \\
            \hline
            32 & 0.0005 & 0.001 & 5 & 73\% & 72\% & 90\% & 0.78 \\
            \hline
            32 & 0.0005 & 0.003 & 9 & 83\% & 80\% & 99\% & 0.63 \\
            \hline
            64 & 0.0005 & 0.003 & 9 & 77\% & 74\% & 97\% & 0.71 \\
            \hline
            64 & 0.003 & 0.03 & 7 & 97\% & 96\% & 100\% & 0.18 \\
            \hline
            128 & 0.008 & 0.001 & 9 & 74\% & 71\% & 95\% & 0.70 \\
            \hline
            128 & 0.0005 & 0.03 & 11 & 99\% & 99\% & 99\% & 0.17 \\
            \hline
            256 & 0.005 & 0.01 & 11 & 97\% & 96\% & 99\% & 0.25 \\
            \hline
            256 & 0.001 & 0.03 & 3 & 99\% & 99\% & 100\% & 0.10 \\
            \hline
        \end{tabular}
    \end{center}
\end{minipage}

\begin{figure}[H]
    \centering
    \includegraphics[width=\textwidth]{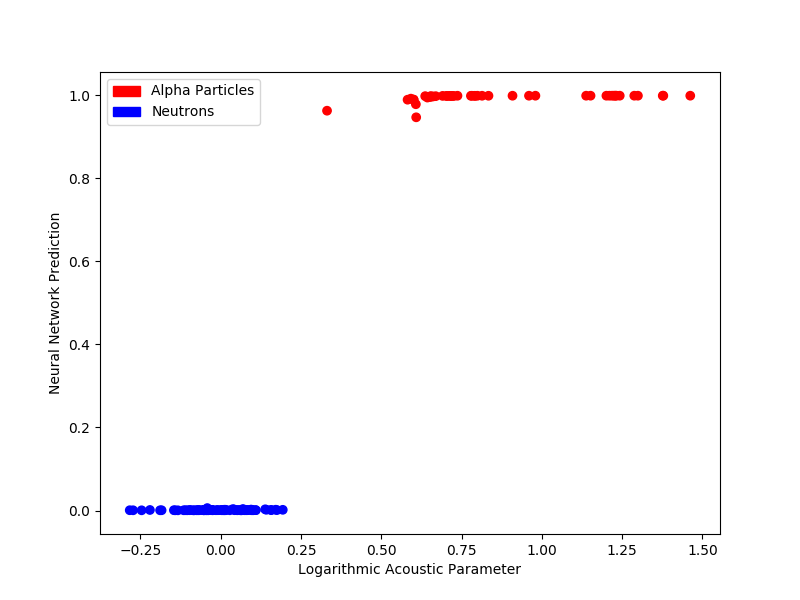}
    \caption{\label{grav_grid_search} Predictions of best gravitational differentiation model compared to AP (on randomly selected validation set).}
\end{figure}

Note that in Figure \ref{grav_grid_search}, AP is normalized using a similar method to the original PICO-60 study \cite{pico}. In this normalized spectrum, the decision boundary between neutrons and alpha particles is at 0.25.

\subsection{Iterative Cluster Nucleation}

\subsubsection{Concept}

Iterative cluster nucleation (ICN) is a second semi-supervised learning algorithm developed as part of this study. It takes advantage of some amount of imperfectly labeled data, using it to classify the rest of an unlabeled training set while optimizing to produce an effective discriminator.

The basic concept is similar to that of gravitational differentiation. Starting with a relatively small amount of labeled data from two classes (nuclear recoils and alpha particles, in this case), predictions are run on a set of unlabeled data, and those predictions are used to produce further training data. However, there are three key differences:

\begin{enumerate}
    \item In iterative cluster nucleation, an unlabeled example is not included in the training process at all, until a very confident prediction is made on it. Once that occurs, it is added to the training set, expanding one of the two ``clusters'' of training data.
    \item Once an example is added to the active training set, it is never removed. This ensures that the highly confident predictions made at the very beginning continue to influence the training process.
    \item The unlabeled examples, once added to the active training set, are weighted just as strongly as examples from the original labeled data set. This ensures imperfect labels in the labeled data do not outweigh the confident and more likely correct predictions from later in the training process.
\end{enumerate}

\subsubsection{Algorithm}

The algorithm is initialized as follows:

\begin{enumerate}
    \item Take a subset of the training data, referred to as the seed set $\varsigma$. Use this as the beginning of the training set.
    \item Remove classifications from any other available training examples. These create the unlabeled set $\upsilon$.
    \item Compile and randomly initialize the weights of a binary classification neural network $NN(x)$.
\end{enumerate}

Parameterized by the initial seed threshold $j$ (on the order of 0.01) and the seed threshold multiplier $k$ (on the order of 1.05), the algorithm follows these iterative steps:

\begin{enumerate}
    \item Train $NN(x)$ for 30 epochs on $\varsigma$, using the imperfect ground truth values available.
    \item Using the partially trained weights of $NN(x)$, run inference on the entirety of $\upsilon$, producing a set of predictions $p$.
    \item Find predictions within $p$ that are within a distance of $j$ of either 0 or 1; as this is a binary classifier, such a prediction represents high confidence. Remove any such examples from $\upsilon$ and add them to $\varsigma$. The principle is that, when $NN(x)$ is trained on a relatively small set for short period of time, the few predictions within $p$ that are very confident are highly likely to be correct.
    \item If no examples have been removed from $\upsilon$ and added to $\varsigma$, multiply $j$ by $k$ in place. This is done to increase the acceptance rate later in the training process, when most easily classifiable examples have been added to $\varsigma$. Otherwise, gridlock would occur, where certain examples could not be confidently classified given the training set.
\end{enumerate}

\subsubsection{Results}

\begin{figure}[h]
    \centering
    \includegraphics[width=\textwidth]{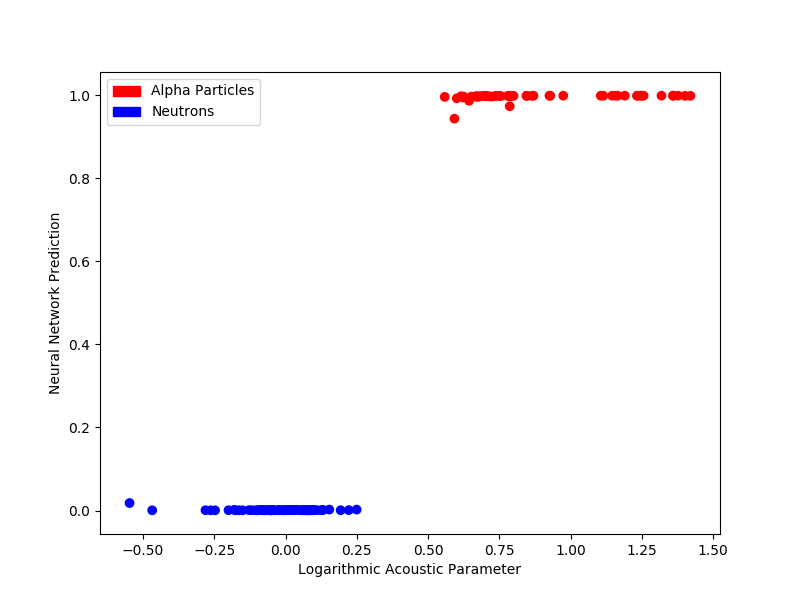}
    \caption{\label{icn_grid_search} Predictions of best iterative cluster nucleation model compared to AP (on randomly selected validation set).}
\end{figure}

During a grid search of these parameters (in addition to regularization parameters of the network), the best mean accuracy over three runs of an individual configuration was 99\%, using 256 initial labeled examples in $\varsigma$. (The network's predictions on the best performing epoch can be observed in Figure \ref{icn_grid_search}.) With 128 examples, the maximum was 98\%; with 64 it was 97\%; and with 32, 95\%. As with gravitational differentiation, larger set sizes improve accuracy significantly.

Once again, the parameters inherent to the ICN algorithm, as well as dropout and L2 regularization, do not have a meaningful polynomial correlation with accuracy. Note that there are several hyperparameters not specifically optimized in the context of iterative cluster nucleation, including the number of layers and numbers of neurons within those layers. The correlation between those hyperparameters and accuracy is subject to future research.

The mean accuracy for a given initial $\varsigma$ is again important. In Table \ref{icn_accuracy_by_varsigma}, it is clear that with small initial $\varsigma$ sets, the training process is very unreliable, much like gravitational differentiation. However, the mean accuracy is much closer to the maximum for initial $\varsigma$ sets of 128 and 256, indicating that the training process of ICN is significantly more consistent than gravitational differentiation.

\begin{minipage}{\textwidth}
    \begin{center}
        \captionof{table}{Iterative cluster nucleation accuracy statistics by initial size of $\varsigma$.} \label{icn_accuracy_by_varsigma}
        \begin{tabular}{|l|l|l|l|}
            \hline
            Initial $\varsigma$ Size & Proportion of Data Labeled & Max Accuracy & Mean Accuracy \\
            \hline
            32 & 6\% & 95\% & 75\% \\
            \hline
            64 & 13\% & 97\% & 84\% \\
            \hline
            128 & 26\% & 98\% & 95\% \\
            \hline
            256 & 52\% & 99\% & 97\% \\
            \hline
        \end{tabular}
    \end{center}
\end{minipage}

Table \ref{icn_overview} represents the performance of a variety of high-accuracy iterative cluster nucleation configurations tested during the grid search. As with gravitational differentiation, the accuracy statistic is the mean over three tests of the same configuration.

\begin{minipage}{\textwidth}
    \begin{center}
        \captionof{table}{Performance of iterative cluster nucleation configurations.} \label{icn_overview}
        \begin{tabular}{|l|l|l|l|l|l|l|l|l|}
            \hline
            Initial $\varsigma$ Size & L2 $\lambda$ & Dropout & Initial $j$ & $k$ & Mean AP Accuracy & Precision & Recall & CWSD \\
            \hline
            32 & 0 & 0.25 & 0.01 & 1.025 & 80\% & 79\% & 99\% & 0.60 \\
            \hline
            32 & 0 & 0 & 0.01 & 1.025 & 95\% & 94\% & 99\% & 0.38 \\
            \hline
            64 & 0.001 & 0 & 0.01 & 1.05 & 97\% & 96\% & 99\% & 0.28 \\
            \hline
            64 & 0 & 0.25 & 0.02 & 1.025 & 97\% & 95\% & 100\% & 0.27 \\
            \hline
            128 & 0 & 0.5 & 0.01 & 1.05 & 92\% & 90\% & 100\% & 0.35 \\
            \hline
            128 & 0 & 0.5 & 0.02 & 1.05 & 98\% & 98\% & 100\% & 0.21 \\
            \hline
            256 & 0 & 0.5 & 0.01 & 1.05 & 99\% & 99\% & 100\% & 0.15 \\
            \hline
            256 & 0.003 & 0.5 & 0.01 & 1.05 & 98\% & 98\% & 98\% & 0.26 \\
            \hline
        \end{tabular}
    \end{center}
\end{minipage}

\subsection{Performance Summary}

Table \ref{semi_success_overview} directly compares performance of the most successful semi-supervised learning configurations, with AP provided as a baseline for comparison. Remember that ``Mean AP Accuracy'' is calculated relative to AP's predictions, rather than replication of the ground truths.

\begin{minipage}{\textwidth}
    \begin{center}
        \captionof{table}{Summary of the most successful semi-supervised learning models.} \label{semi_success_overview}
        \begin{tabular}{|l|l|l|l|l|l|}
            \hline
            Initial $\varsigma$ Size & Technique & Mean AP Accuracy & Mean Precision & Mean Recall & Mean CWSD \\
            \hline
            32 & GD & 83\% & 80\% & 99\% & 0.63 \\
            \hline
            32 & ICN & 95\% & 94\% & 99\% & 0.38 \\
            \hline
            64 & GD & 97\% & 96\% & 100\% & 0.18 \\
            \hline
            64 & ICN & 97\% & 96\% & 99\% & 0.28 \\
            \hline
            128 & GD & 99\% & 99\% & 99\% & 0.17 \\
            \hline
            128 & ICN & 98\% & 98\% & 100\% & 0.21 \\
            \hline
            256 & GD & 99\% & 99\% & 100\% & 0.10 \\
            \hline
            256 & ICN & 99\% & 99\% & 100\% & 0.15 \\
            \hline
            278 & PICO-60 ML & 80\% & 76\% & 100\% & 0.64 \\
            \hline
            N/A & AP & 100\% & 100\% & 100\% & 0.41 \\
            \hline
        \end{tabular}
    \end{center}
\end{minipage}

As observed in this table, semi-supervised learning algorithms are significantly more accurate at replicating AP than the original PICO-60 machine learning analysis. Compared to AP, in addition to achieving similar accuracy, configurations from both semi-supervised learning algorithms produce significantly more decisive predictions as measured by the class-wise standard deviation.

Table \ref{average_semi} compares the overall average performance of each of the semi-supervised learning configurations for different labeled set sizes.

\begin{minipage}{\textwidth}
    \begin{center}
        \captionof{table}{Summary of the average accuracy values for each semi-supervised learning technique.} \label{average_semi}
        \begin{tabular}{|l|l|l|}
            \hline
            Initial $\varsigma$ Size & Technique & Mean Accuracy \\
            \hline
            32 & GD & 69\% \\
            \hline
            64 & GD & 73\% \\
            \hline
            128 & GD & 84\% \\
            \hline
            256 & GD & 94\% \\
            \hline
            32 & ICN & 75\% \\
            \hline
            64 & ICN & 84\% \\
            \hline
            128 & ICN & 95\% \\
            \hline
            256 & ICN & 97\% \\
            \hline
        \end{tabular}
    \end{center}
\end{minipage}

\section{Conclusions}

This study has demonstrated that it is possible, using semi-supervised learning, to automatically optimize a discriminator with performance comparable to one developed and tuned by humans like the Acoustic Parameter technique. This conclusion has been achieved through two rounds of optimization, the first to explore and evaluate various network structures and input formats using supervised learning, and the second to incorporate the most successful configuration into semi-supervised learning systems with the goal of producing a high-accuracy discriminator to complement AP.

In the realm of supervised learning, it was made clear that the banded Fourier transform is the most effective input format. It produced significantly higher accuracy and tighter prediction distributions than neural networks trained on either the raw audio waveform or image window data. There is strong evidence that image data in particular does not contain sufficient information to be used for discrimination.

Two semi-supervised learning algorithms were developed, applying the best configuration found for supervised learning. While both techniques produced some very effective discriminators, iterative cluster nucleation was found to be the more effective of the two. As seen in Table \ref{average_semi}, it replicated AP with a mean of 97\% accuracy, using 256 labeled training examples. This indicates that the technique shows significant promise.

As discussed in Section \ref{objective}, discriminators such as AP are difficult to develop until a large quantity of calibration data has been collected and the physical principles of the experiment are understood well. The innovations developed in this study suggest that semi-supervised learning can efficiently replicate AP, thus facilitating very quick development of discriminators immediately after the first calibration runs are conducted on a new experimental apparatus. This has the potential to allow the teams working on these experiments to iterate more quickly, and to better understand intermediate results.

\section{Technologies}

All programming for this study was done in Python 3. Keras \cite{keras}, running on a TensorFlow \cite{tensorflow} backend, was used for all machine learning tasks. NumPy \cite{numpy} and SciPy \cite{scipy} were used for linear algebra and signal processing. ROOT \cite{root}, scikit-image \cite{scikit-image}, and scikit-learn \cite{scikit-learn} were used for data loading and storage. Matplotlib \cite{matplotlib} was used for data visualization.

Implementation code for all algorithms discussed in this paper can be found at \url{https://github.com/brendon-ai/dark-matter}.

\section{Acknowledgements}

The PICO collaboration wishes to thank SNOLAB and its staff for support through underground space, logistical and technical services. SNOLAB operations are supported by the Canada Foundation for Innovation and the Province of Ontario Ministry of Research and Innovation, with underground access provided by Vale at the Creighton mine site. We wish to acknowledge the support of the Natural Sciences and Engineering Research Council of Canada (NSERC) and the Canada Foundation for Innovation (CFI) for funding. We acknowledge the support from National Science Foundation (NSF) (Grant 0919526, 1506337, 1242637 and 1205987). We acknowledge that this work is supported by the U.S. Department of Energy (DOE) Office of Science, Office of High Energy Physics (under award DE-SC-0012161), by the DOE Office of Science Graduate Student Research (SCGSR) award,  by DGAPA-UNAM (PAPIIT No.\:IA100316 and IA100118) and Consejo Nacional de Ciencia y Tecnolog\'ia (CONACyT, M\'exico, Grant No.\:252167), by the Department of Atomic Energy (DAE), the Government of India, under the Center of AstroParticle Physics II project (,CAPP-II) at SAHA Institute of Nuclear Physics (SINP), European Regional Development Fund-Project ``Engineering applications of microworld physics'' (No.\:CZ.02.1.01/0.0/0.0/16\_019/0000766), 
and the Spanish Ministerio de Econom\'ia y Competitividad, Consolider MultiDark (Grant CSD2009-00064). This work is partially supported by the Kavli Institute for Cosmological Physics at the University of Chicago through NSF grant 1125897, and an endowment from the Kavli Foundation and its founder Fred Kavli. We also wish to acknowledge the support from Fermi National Accelerator Laboratory under Contract No.\:DE-AC02-07CH11359, and Pacific Northwest National Laboratory, which is operated by Battelle for the U.S. Department of Energy under Contract No.\:DE-AC05-76RL01830. We also thank Compute Canada (\url{www.computecanada.ca}) and the Centre for Advanced Computing, ACENET, Calcul Qu\'ebec, Compute Ontario and WestGrid for the computational support.

\printbibliography

\end{document}